\documentstyle[prl,aps,twocolumn,epsf]{revtex}
\def\break#1{\pagebreak \vspace*{#1}}
\begin{document}

\draft

\title{ Supersymmetric Fokker-Planck strict isospectrality }

\author{Haret C. Rosu
\cite{byline}
}

\address{ 
{Instituto de F\'{\i}sica de la Universidad de Guanajuato, Apdo Postal
E-143, Le\'on, Gto, M\'exico}\\
{Institute of Gravitation and Space Sciences, P.O. Box MG-6,
Magurele-Bucharest, Romania}\\
 }

\maketitle
\widetext

\begin{abstract}

I report a study of the nonstationary one-dimensional Fokker-Planck solutions
by means of the strictly isospectral method of supesymmetric
quantum mechanics. The main
conclusion is that this technique can lead to a space-dependent
(modulational) damping of the spatial part of the nonstationary
Fokker-Planck solutions, which I call strictly isospectral damping.
At the same time, using an additive decomposition of the nonstationary
solutions suggested by the strictly isospectral procedure and by an
argument of Englefield [J. Stat. Phys. {\bf 52}, 369 (1988)], they can be
normalized and thus turned into physical solutions, i.e., Fokker-Planck
probability densities. There might be applications to many
physical processes during their transient period.

\end{abstract}


\narrowtext


At present, supersymmetry is recognized as a basic symmetry of natural world
being discussed with various degrees of sophistication in many research areas.
We also know that in physics it happens quite often that simple mathematical
procedures might
be extremely efficient and may lead to significant progress in clarifying
our vision of the world. This is undoubtedly the case of Witten's
supersymmetric quantum mechanics \cite{w81} which, mathematically speaking, is
the factorization of the one dimensional (1D)
Schr\"odinger operator and relies on the Darboux transformations.
The procedure involves Riccati equations for the so called superpotential
and most people got used to work with the
particular Riccati solution. However, Mielnik \cite{m84} realized that one
can use the general Riccati solution and presented the 1D
harmonic oscillator in that perspective, hinting on the connection with the
inverse scattering methods. A clear discussion was provided by Nieto
\cite{n84} and later Sukumar \cite{s85} showed that the Gel'fand-Levitan method
can be interpreted as a sequence of two Darboux transformations.
While the particular superpotential leads to the
quasi-isospectrality of the partner Hamiltonians, the general one allows for
the introduction of entire families of
strictly  isospectral of either ``boson" or ``fermion" partner systems.
In this paper I shall call Darboux-Witten (DW) isospectrality
this strict isospectrality, which is obtained by employing the
general superpotential (other people call it the double Darboux method).
In a recent paper, Sukhatme and
collaborators \cite{psp93} obtained bound states in the
continuum by means of DW isospectrality and a
couple of other applications may be found in the literature \cite{r96}.
In the following,
after quickly presenting the factorization of the 1D Fokker-Planck (FP)
equation, I shall tackle its DW (strict) isospectrality.
Before proceeding, I recall that Darboux transformations (not DW ones)
have been used
by Englefield \cite{e88} for a FP equation with a particular potential,
while Hron and Razavy \cite{hr85} studied some solvable models of the
\break{1.62in}
FP equation by means of the Gel'fand-Levitan method.
The other two techniques similar to the DW isospectrality,
namely, the Abraham-Moses procedure \cite{am80}
and the Pursey one \cite{p86} can be applied to the FP equation without
any difficulty.

Bernstein and Brown \cite{bb84}
were the first people to provide a simple discussion of the
correspondence between the 1D FP equation with an
arbitrary potential and Witten's supersymmetric quantum mechanics.
They showed that
the great advantage of the supersymmetric procedure for the FP problem
is to replace bistable ``bosonic" potentials with much simpler
single well ``fermionic" ones.
Here, I shall use the Smoluchowski form of the 1D FP equation with constant
diffusion coefficient and potential drift
\begin{equation} \label{Sm}
\frac{\partial}{\partial t}{\cal P }(x,t)=
\frac{\partial}{\partial x}
\Bigg[\frac{\partial}{\partial x}+\gamma
f'(x)\Bigg]{\cal P}(x,t)
\end{equation}
where $f'=df/dx$ is the drift force up to a sign, i.e., $f$ is the FP
potential, and $\gamma$ is a free
parameter whose expression depends on the physical problem under consideration.
Eq.~(1) is not a
Hamiltonian evolution of the solutions. Nevertheless it can be cast
into the Schr\"odinger equation as follows.
Any initial time-dependent solution ${\cal P}(x,t)$ will relax at asymptotic
times to the stationary solution

\begin{equation} \label{P1}
{\cal P}_{0}(x)=
\exp [-\gamma f(x)]~,
\end{equation}
but to pass to probability densities ${\bf P}$ one should introduce
a normalization constant $N^{-1}_{C}$ as a factor in the right hand side
of Eq. (2).
The issue of normalization looks obvious but let me emphasize that to make it
clear one should consider the FP probability current.
This important quantity occurs when one wants to turn the FP equation into a
continuity one. It reads
\begin{equation} \label{J}
J(x,t)=-e^{-\gamma f}[e^{\gamma f}{\bf P}]'~.
\end{equation}
The stationarity is defined as the case(s) of constant $J$, and the constant
is determined by the boundary conditions. The particular case of zero $J$
means no periodic boundary conditions and leads to
$e^{\gamma f}{\bf P}_{0}=C$, i.e., the constant is $C=1/N_{C}$.
To get the common notion of
probabilities (a set of positive numbers, each smaller than unity,
summing up to unity,
and mapped on an equivalent set of future, possible events belonging to the
same class)
one should take the normalization constant as follows
\begin{equation}  \label{N}
N_{C}=\int e^{-\gamma f}dx~,
\end{equation}
where the integration limits are from zero to $\infty$ and from
$-\infty$ to $\infty$ for the half-line and the full line cases, respectively.

The evolution at intermediate
times can be discussed conveniently by means of the celebrated ansatz
\begin{equation} \label{P2}
{\cal P}(x,t)= \psi (x,t) \exp \left(-\frac{1}{2}\gamma f(x)\right)
=
\psi (x,t)\sqrt{{\cal P}_0(x)}~.
\end{equation}
${\cal P}(x,t)\rightarrow {\cal P}_{0}$
when $t\rightarrow \infty$, that is $\psi (x,t)
\propto \sqrt{{\cal P}_0(x)}$ at asymptotic times. The ansatz (5) 
turns the FP evolution of ${\cal P}$ into
a Schr\"odinger evolution for the amplitude function $\psi$ in
imaginary time 
\begin{equation}  \label{P22}
\frac{\partial \psi}{\partial t}=-H_{FP}\psi~,
\end{equation}
where the FP Hamiltonian is a Hermitian and positive semidefinite operator.
It is now easy to proceed with the factorization and the
whole Witten scheme for the FP Hamiltonian. We write
$ H_{FP,1}=A_1A_2$,
with
$A_1=e^{\gamma f/2}\frac{\partial}{\partial x}
e^{-\gamma f/2}
\equiv
\frac{\partial}{\partial x} -\frac{1}{2} \gamma f'$
and
$A_2=e^{-\gamma f/2}\frac{\partial}{\partial x}
e^{\gamma f/2}
\equiv\frac{\partial}{\partial x} + \frac{1}{2}\gamma f'$. I recall that the
usual quantum-mechanical factorization is of the sort 
$H_{qm}=AA^{\dagger}+\epsilon$, where $\epsilon$ is the so-called factorization
energy, which is a parameter fixing the energy scale. A zero value of this 
parameter, as in the case of the FP equation, indicates that the factoring of 
the Hamiltonian is done with respect to the ``ground state" wavefunction. One 
can easily see that the FP superpotential is 
${\cal W} _{0}=\frac{d}{dx}\ln\sqrt{{\cal P}_{0}}$ and thus the mentioned FP
factorization is indeed with respect to the FP ground state amplitude.
Also, I remark that the FP superpotential
${\cal W} _{0}=-\frac{1}{2}\gamma f'$ is
proportional to the drift force, which is a quite well-known result.
The superpartner Hamiltonian will be $H_{FP,2}= A_2A_1$. The two FP
Hamiltonian partners can be written as follows
\begin{equation} \label{HFP}
-H_{FP,1,2}=\frac{d^2}{dx^2}-S_{1,2}
\end{equation}
with the Schr\"odinger potentials $S_{1,2}$ entering simple Riccati equations
\begin{equation} \label{S1}
S_{1,2}=e^{\pm\gamma f/2}\left(\frac{d^2}{dx^2}e^{\mp\gamma f/2}
\right)=\left(\frac{\gamma f'}{2}\right)^{2}\mp \frac{\gamma f''}{2}.
\end{equation}

At this point, it is worthwhile to emphasize the connection between the FP
solutions and the Schr\"odinger solutions. According to Eq.~(5), the FP
solutions can always be decomposed into a product of a non-stationary
Schr\"odinger solution and the ground state amplitude of the corresponding
stationary Schr\"odinger equation.
As remarked by Englefield \cite{e88}, $\sqrt{{\cal P} _{0}}$ in Eq.~(5) can be
any positive solution of the stationary Schr\"odinger equation, i.e., does
not have to be normalizable. This is a crucial point for our arguments below.

Let us pass now to the strictly isospectral construction.
In principle, one can use any solution $\varphi (x)$ of the {\em stationary}
Schr\"odinger equation to perform this construction \cite{psp93,r96}, but we
shall
use the ground state amplitude $\varphi _{0}(x) =\sqrt{{\cal P} _{0}}$. It
is saying that the strictly isospectral Schr\"odinger potentials are given by
\begin{eqnarray} \nonumber
S^{DW}_{iso,1}(x, \varphi _{0} ;\lambda) &=&
S_1(x)-2[\ln({\cal J} +\lambda)]''\\ \label{S3}
&=&
S_1(x)-\frac{4\varphi _{0}\varphi _{0}{'}}{{\cal J} +\lambda}+
\frac{2\varphi _{0}^{4}}{({\cal J} +\lambda)^2}~,
\end{eqnarray}
where $\lambda$ is a real parameter, that mathematically is the arbitrary
integration constant of the general Riccati solution, and
\begin{equation} \label{CJ}
{\cal J} (x)\equiv\int_{c}^{x}\varphi _{0}^{2}(y)dy~,
\end{equation}
where $c=0$ for the half line case and $c=-\infty$ for the full line one,
whereas the expression of $\varphi _{gen}$ is given below.
In this construction
the $\varphi _{0}$ solution is reintroduced in the spectrum by
using the general superpotential solution of the Riccati
equation which reads
\begin{equation} \label{wgen}
{\cal W} _{gen}= -\frac{1}{2}\gamma f^{'}+      
 \frac{d}{dx}\ln [{\cal J} (x)+\lambda]~.
\end{equation}
The way to obtain Eq.~(\ref{wgen}) is well-known \cite{m84,n84} and will not
be repeated here.
From it one can easily get Eq.~(9). The new class of solutions
is a one-parameter family
$\varphi _{gen}$ differing from $\varphi _{0}$, by a quotient
\begin{equation} \label{var}
\varphi _{gen}(x;\lambda)=
\frac{\varphi _{0}(x)}{{\cal J} +\lambda}=\varphi _{0}(x)/M_{D}(x)~.
\end{equation}
In the FP context we need $\varphi _{gen}>0$, i.e.,
$\lambda \in (-{\cal I}(x),\infty)$. The range of $\lambda$ will be further
restricted by a normalization condition (see below).
The main point now is that the denominator $M_{D}(x)$ looks like a
space-dependent
modulation. Its general behavior has been obtained in \cite{psp93,r96}.
There is a strong damping effect of the integral ${\cal J}$,
both for the
family of potentials as for the wavefunctions. The parameter $\lambda$ is
just signaling the importance of the integral term. There is some modulation
close to the origin and for small $\lambda$ parameters. At higher
$\lambda$-values the damping nature gets extremely strong and $\varphi _{gen}$
is going rapidly to zero.

The general form of the FP amplitude solution using the general
Riccati solution reads
\begin{equation} \label{Pgen}
\varphi_{gen}=\exp\left(\int {\cal W}_{gen}\right)= \exp(-\frac{1}{2}
\gamma f)/M_{D}
\end{equation}
and thus one can see the self-modulation of the solutions.
These modulated solutions are stationary solutions of the
following Schr\"odinger equation
\begin{equation} \label{NLS}
\frac{d^2\varphi (x)}{dx^2}-S^{DW}_{iso,1}(x, \varphi _{0} ;\lambda)
\varphi (x)=0~,
\end{equation}
corresponding to FP equations which
are strictly isospectral to the initial one. Exactly as in supersymmetric
quantum mechanics, the strictly isospectral FP equations
can be understood as a one-parameter family of ``bosonic"
FP equations having the same ``fermionic" partner equation.
Its factoring operators are $B_1=\frac{\partial}{\partial x}+
{\cal W}_{gen}$ and $B_2=\frac{\partial}{\partial x}-{\cal W}_{gen}$.
The amplitude $\psi _{1}=B_{1}\psi$ satisfies the time-dependent
Schr\"odinger equation Eq.~(6) unless $S_{2}$
substitutes $S_{1}$. Using now the essential fact that $1/\varphi _{gen}$
(``ground state zero-mode") is
the general solution of Eq.~(6) again when $S_2$ replaces $S_1$ (see section
7.1 in the review of Cooper, Khare and Sukhatme \cite{w81}), and being
positive can be employed in Eq.~(5) instead of
$\sqrt{{\cal P} _{0}}$ whenever $\psi _{1}$ substitutes $\psi$, one gets
the following FP solution
\begin{equation} \label{new}
{\cal P} _{1}(x,t)=\varphi _{gen}^{-1}\cdot(B_{1}\psi)=
\frac{\partial}{\partial x}\left(\frac{\psi}{\varphi _{gen}}\right)~.
\end{equation}
If one calculates the normalization integral for this solution one gets
\begin{equation} \label{norm}
N _1=\int _{-\infty}^{+\infty}{\cal P} _{1}(x,t)dx=
(\psi\varphi _{gen}^{-1})|_{-\infty}^{+\infty}~.
\end{equation}
For confining FP potentials, i.e., going to $\infty$ at both asymptotic
limits one has $\frac{1}{\varphi _{gen}}\rightarrow 0$ for any $\lambda$
and therefore $N_{1}$ is zero.
On the other hand, $\varphi _{gen}$ can be normalized and of course
adding any multiple of ${\cal P} _{1}$ will not change the
normalization constant.
In this way, one arrives at strictly isospectral
FP probabilities of the type
\begin{equation} \label{FP2}
{\bf P} _{2}(x,t)= k{\cal P} _{1}(x,t)+ N_{gen}^{2}\varphi _{gen}^{2}~,
\end{equation}
where the normalization constant $N_{gen}$ is \cite{priv}
\begin{equation}  \label{N2}
N_{gen}=
\int _{-\infty}^{+\infty}\varphi _{gen}^{2}(x)dx=\sqrt{\lambda(\lambda+1)}~.
\end{equation}
The constant $k$ is
arbitrary, restricted by the physical condition ${\bf P} _{2}\geq 0$.
Moreover, $N_{gen}$ imposes further restrictions on the $\lambda$ parameter.
The correct range of $\lambda$ in any physical problem related to FP
transients should be $\lambda \in (-{\cal I} (x), -1)\cup (0,\infty)$.

The decomposition Eq.~(17) of the transient FP probabilities appears to
be quite general. The behavior of the transient probabilities is dictated
by the parameter $\lambda$, which is also essential in the normalization
issue. On the other hand, since $S_{iso}^{DW}\rightarrow S_{1}$ for
$\lambda \rightarrow \infty$, the decomposition Eq.~(17) does not apply
any more in that limit, and one should use the common decomposition
Eq.~(5). In other words,
although the isospectral solutions are stationary ones they are related
to the transient period and disappear in the asymptotic stationary
regime. The parameter $\lambda$ looks like an effective ``time" (damping)
parameter measuring their importance during the transient phases.


Perhaps it is worthwhile to remark that in our notations
the initial FP equation reads
\begin{equation} \label{Sm1}
\frac{\partial}{\partial t}{\cal P }(x,t)=
\frac{\partial}{\partial x}
\Bigg[A_2+\frac{1}{2}\gamma f^{'}\Bigg]{\cal P}(x,t)~.
\end{equation}
Substituting $A_2$ by $A_1$ gives a FP form of the heat equation
\begin{equation} \label{H}
\frac{\partial}{\partial t}{\cal P }(x,t)=
\frac{\partial}{\partial x}
\Bigg[A_1+\frac{1}{2}\gamma f^{'}\Bigg]{\cal P}(x,t)~,
\end{equation}
since the operator in the brackets is $\partial/\partial x$.

The strictly isospectral FP equations can be written by analogy as follows
\begin{equation} \label {IFP1}
\frac{\partial}{\partial t}{\cal P } (x,t)=
\frac{\partial}{\partial x}
\Bigg[B_2+\frac{1}{2}\gamma f^{'}\Bigg]{\cal P} (x,t)
\end{equation}
and the corresponding isospectral ``heat" equation will be
\begin{equation} \label {IFP2}
\frac{\partial}{\partial t}{\cal P } (x,t)=
\frac{\partial}{\partial x}
\Bigg[B_1+\frac{1}{2}\gamma f^{'}\Bigg]{\cal P} (x,t)~.
\end{equation}

Finally, I recall that besides the DW strictly isospectral potential
Eq.~(9), there are four other distinct families of strictly isospectral
Schr\"odinger potentials obtained from pair combinations of
Darboux-Witten, Pursey, and Abraham-Moses procedures \cite{ks}.
The formulas corresponding to Eq.~(9) for the four families
are collected in the paper
of Khare and Sukhatme \cite{ks}. All of them are of the Darboux type and
one can use any of them in the FP problem.
From the scattering point of view all these families have the same
reflection and transmission probabilities. However, they differ from each
other in reflection and transmission amplitudes.


In conclusion, I have shown in a simple and general manner that
DW strict isospectrality
may provide interesting insights in the case of FP probabilities.

{\bf Acknowledgment}

This work was supported in part by the Consejo Nacional de Ciencia y
Tecnolog\'{\i}a (Mexico) Project 4868-E9406.

\end{document}